# A Minimal Physics-Based Model on the Electrochemical Impedance Spectroscopy of Solid-State Electrolyte

Changyu Deng[a], and Wei Lu[a,b]

[a] Department of Mechanical Engineering, University of Michigan, Ann Arbor, Michigan 48109, USA
[b] Department of Materials Science & Engineering, University of Michigan, Ann Arbor, Michigan 48109, USA

Solid state batteries have emerged as a potential next-generation energy storage device due to safety and energy density advantages. Development of electrolyte is one of the most important topics in solid state batteries. Electrochemical Impedance Spectroscopy (EIS) is a popular measurement technique to obtain the conductivity and diagnose the electrolyte. Current interpretation mainly uses the semicircle part of the curves and discards other information revealed by EIS such as the slope of the curve at low frequency. What is worse, some features on the curve are not fully interpreted. To better understand the transport mechanism and interpret EIS curves, we introduce a continuous model to quantify the ion transport and current flow in the electrolyte. The produced EIS curves from the model are compared with experiment data to show good agreement.

## 1. Introduction

Lithium and sodium ion batteries are essential energy storage devices nowadays [1–3]. Solid state batteries have been attracting much attention due to safety and energy density advantages over traditional liquid electrolyte cells [4]. Electrochemical Impedance Spectroscopy (EIS), which applies an alternating current and calculate impedance as a function of current frequency, is a popular technique in battery analysis [5]. It is applied to solid state electrolyte to measure ionic conductivity and characterize chemical reactions [6,7].

Despite the wide use of EIS, the post-processing often oversimplifies the transport mechanism in the solid-state electrolyte. Equivalent circuit models, composed of resistors, capacitors and so forth, are mostly used to fit and interpret EIS curves[8]. As a result, only the intersection of the EIS curve with the real axis matters. This simple interpretation, however, discards other information revealed by EIS such as the slope of the curve at low frequency. What is worse, some features on the curve are not fully interpreted. For instance, the semicircle in the Nyquist plot, often regarded as charge transfer, also occurs when the electrolyte is connected to two non-reacting electrodes such as stainless steel and gold [9]. On the other hand, existing physics-based models are mostly microscale [10,11]. Besides the complexity to leverage them in EIS, there are many undetermined parameters in the model which require additional experiments.

To better understand the transport mechanism in solid state electrolyte and interpret EIS curves, we introduce a continuous model to quantify the ion transport and current flow in the electrolyte. We aim to use minimal physics-based equations to build the model so that the model parameters can be easily correlated with the measured curves. We verify the model by experiments. Ablation study and sensitivity analysis are conducted.

## 2. Experimental

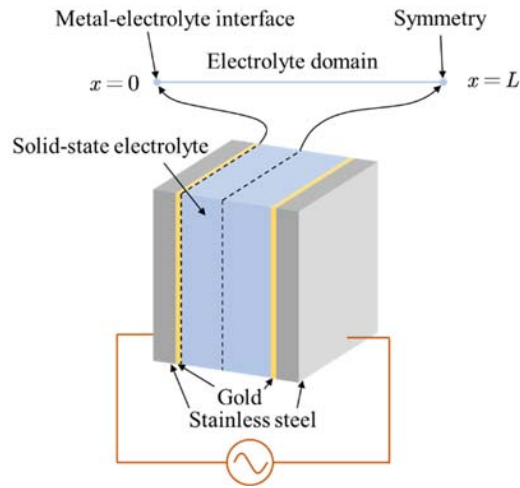

Figure 1 Illustration of experimental setup and model domain. Gold is coated on electrolyte surface. The stainless steel current collector is under an external force to make good contact with gold. Because of symmetry in the configuration, we only need to consider half of the system (i.e. from $x=0$ to the symmetry line of $x=L$).

We first introduce the measurement setup of EIS, which follow a typical simple setup. We use a Li$_2$Al$_2$SiP$_2$TiO$_{13}$ (LASPT) ion conductive ceramic sheet (from MTI) as the solid-state electrolyte material, whose dimensions are 5 mm × 5 mm × 0.16 mm. As shown in Figure 1, gold layers are coated on the two square surfaces of the material by a sputter machine to reduce contact resistance. Two stainless steel films are squeezed on the left and right side, respectively. A sinusoidal current, whose amplitude is 100 mV, is applied to the stainless steel.

## 3. Model

For simplicity, we elaborate our model by a one-dimensional case in Figure 1. The model can be easily extended to higher dimensions.

### 3.1 Time domain

Within the electrolyte domain, we consider the transport of lithium ions and electrons. They both follow mass conservation

$$\frac{\partial c_1}{\partial t} + \frac{\partial j_1}{\partial x} = 0, \tag{1}$$

$$\frac{\partial c_2}{\partial t} + \frac{\partial j_2}{\partial x} = 0, \tag{2}$$

where subscript 1 denotes lithium ions, subscript 2 denotes electrons, $c_i (i=1,2)$ denotes concentration, $j_i$ denotes mass flux, $t$ denotes time. Net charges will contribute to the potential by

$$-\varepsilon \frac{\partial^2 \phi}{\partial x^2} = F(c_1 - c_2) \tag{3}$$

where $\varepsilon$ denotes permittivity, $\phi$ denotes potential, $F$ denotes the Faraday constant. Flux of two charge carriers is caused by diffusion and electrophoresis,

$$j_1 = -D_1 \frac{\partial c_1}{\partial x} - u_1 F c_1 \frac{\partial \phi}{\partial x}, \tag{4}$$

$$j_2 = -D_2 \frac{\partial c_2}{\partial x} + u_2 F c_2 \frac{\partial \phi}{\partial x}, \tag{5}$$

where $D_i$ denotes diffusivity and $u_i$ denotes mobility. Note that the signs before $u_i$ are different due to the signs of charges.

At *x*=0, the potential is equal to the applied voltage

$$\phi|_{x=0} = \phi_{ext}, \tag{6}$$

where $\phi_{ext}$ denotes external potential (from the alternating current). Lithium ions cannot pass through the metal surface, giving

$$j_1\big|_{x=0}=0. \tag{7}$$

Therefore, electron is the only charge carrier across the interface between the solid electrolyte and the metal. Although the double layer near the boundary will build up an electric field at the interface and accordingly induce current when its intensity changes, we assume the induced current is negligible compared with the current carried by the electron charge transfer. Namely, we assume zero electric field at the boundary,

$$\frac{\partial \phi}{\partial x}\bigg|_{x=0}=0. \tag{8}$$

At $x=L$, we set the potential reference

$$\phi\big|_{x=L}=0. \tag{9}$$

Due to symmetry, the concentration of both charge carriers keep constant in an alternating current,

$$c_1\big|_{x=L}=c_0, \tag{10}$$

$$c_2\big|_{x=L}=c_0, \tag{11}$$

where $c_0$ is the initial equilibrium concentration.

### 3.2 Frequency domain

The applied voltage can be written as

$$\phi_{ext}=Ve^{i\omega t}, \tag{12}$$

where $V$ is the amplitude of the applied voltage and $\omega$ is frequency. Therefore, it is convenient to convert all equations to the frequency domain. Then the variables are only functions of coordinate $x$, so that the above equations become ordinary differential equations. We can rewrite the governing equations as

$$i\omega c_1 + j_1' = 0, \tag{13}$$

$$i\omega c_2 + j_2' = 0, \tag{14}$$

$$-\varepsilon\phi'' = F(c_1 - c_2), \tag{15}$$

$$j_1 = -D_1 c_1' - u_1 F c_1 \phi', \tag{16}$$

$$j_2 = -D_2 c_2' + u_2 F c_2 \phi'. \tag{17}$$

At $x=0$ we have

$$j_1\big|_{x=0}=0, \tag{18}$$

$$\phi'|_{x=0}=0. \tag{19}$$

At $x=L$, we have

$$\phi|_{x=L}=0, \tag{20}$$

$$c_1|_{x=L}=c_0, \tag{21}$$

$$c_2|_{x=L}=c_0. \tag{22}$$

### 3.3 Dimensionless simplification

The frequency domain equations are non-linear differential equations, because conductivity is proportional to $c_1$ in Eq. (16) or $c_2$ in Eq. (17). Since the voltage $V$ is small, we can ignore the change of conductivity and replace $c_1$ and $c_2$ by $c_0$. Besides, we can nondimensionalize the equations by $\tilde{x}=x/L$, $\tilde{c_i}=\dfrac{FL^2(c_i-c_0)}{\varepsilon V}$, $\tilde{\omega}=\dfrac{L^2\omega}{D_1}$, $\tilde{\phi}=\dfrac{\phi}{V}$, $\tilde{j_i}=\dfrac{L^3 F}{\varepsilon V D_i}j_i$, $\tilde{u_i}=\dfrac{F^2 L^2 c_0 u_i}{D_i \varepsilon}$, and $\tilde{D}=\dfrac{D_1}{D_2}$. The domain equations become

$$i\tilde{\omega}\tilde{c_1}+\tilde{j_1}'=0, \tag{23}$$

$$i\tilde{D}\tilde{\omega}\tilde{c_2}+\tilde{j_2}'=0, \tag{24}$$

$$-\tilde{\phi}''=\tilde{c_1}-\tilde{c_2}, \tag{25}$$

$$\tilde{j_1}=-\tilde{c_1}'-\tilde{u_1}\tilde{\phi}', \tag{26}$$

$$\tilde{j_2}=-\tilde{c_2}'+\tilde{u_2}\tilde{\phi}'. \tag{27}$$

The boundary conditions at $x=0$ become

$$\tilde{\phi}|_{x=0}=1, \tag{28}$$

$$\tilde{j_1}|_{x=0}=0. \tag{29}$$

The boundary conditions at $x=L$ become

$$\tilde{\phi}|_{x=1}=\tilde{c_1}|_{x=1}=\tilde{c_2}|_{x=1}=0. \tag{30}$$

The above equations have an analytical solution, although the formula is too long to list here.

### 4. Results

It is widely shown that diffusivity is a function of frequency [12]. Even if local diffusivity does not change with frequency, effective overall diffusivity will change with frequency due to spatial variance [13]. Therefore, we express the diffusivity by a form of

$$D_i(\omega) = \frac{a + bi + i\omega^\gamma}{n(a + bi) + i\omega^\gamma} \tag{31}$$

where $\gamma \in [0, 1]$ and $n \geq 1$ (which means diffusivity increases with frequency).

We use the EIS data in the frequency range of $10^0 \sim 10^{5.5}$ Hz to fit the data. We select this frequency range since both higher and lower frequency increases measurement uncertainty due to noise. The results show good agreement between our model and experiment.

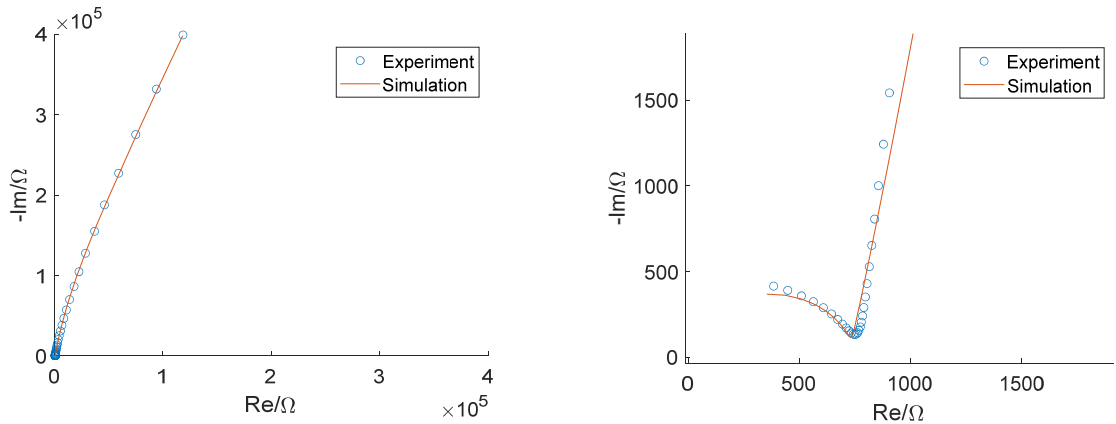

Figure 2 Comparison between simulation and experiment data. Right figure is an enlarged view of the left figure.

## 5. Conclusion

This paper presents a physics-based continuum model to model the solid-state electrolyte under EIS measurement. We develop a minimal physics-based model which provides direct understanding of the EIS response. The model shows good agreement with experimental data.